\documentclass[12pt]{article}

\newsavebox{\ns}
\newsavebox{\dbrane}
\newsavebox{\dbshort}

\usepackage{epsfig}
\usepackage{amssymb}

\def\be{\begin{eqnarray}}
\def\ee{\end{eqnarray}}

\newcommand{\nn}{\nonumber}
\newcommand\para{\paragraph{}}
\newcommand{\ft}[2]{{\textstyle\frac{#1}{#2}}}
\newcommand{\eqn}[1]{(\ref{#1})}

\def\Dslash{\,\,{\raise.15ex\hbox{/}\mkern-12mu D}}
\def\Dbarslash{\,\,{\raise.15ex\hbox{/}\mkern-12mu {\bar D}}}
\def\delslash{\,\,{\raise.15ex\hbox{/}\mkern-9mu \partial}}
\def\delbarslash{\,\,{\raise.15ex\hbox{/}\mkern-9mu {\bar\partial}}}
\def\pslash{\,\,{\raise.15ex\hbox{/}\mkern-9mu p}}
\def\calDslash{\,\,{\raise.15ex\hbox{/}\mkern-12mu {\cal D}}}

\def\be{\begin{eqnarray}}
\def\ee{\end{eqnarray}}

\def\nn{\nonumber}
\def\Tr{\rm{Tr}}

\begin{document}
\pagestyle{plain}
\setcounter{page}{1}
\newcounter{bean}
\baselineskip16pt

\begin{titlepage}

\begin{center}
\today
\hfill {.}\\
%\hfill MIT-CTP-3390 \\

\vskip 1.5 cm {\large \bf Membranes on an Orbifold}
\vskip 1cm
{Neil Lambert${}^1$ and David Tong${}^2$}\\
\vskip 1cm {\sl ${}^1$Department of Mathematics, \\ Kings College London, UK}
\vskip 0.2cm {\sl ${}^2$Department of Applied Mathematics and Theoretical
Physics, \\ University of Cambridge, UK}

\end{center}

\vskip 0.5 cm
\begin{abstract}
We harvest clues to aid with the interpretation of the recently
discovered ${\cal N}=8$ supersymmetric Chern-Simons theory with
$SO(4)$ gauge symmetry. The theory is argued to describe two membranes
moving in the orbifold ${\bf R}^8/{\bf Z}_2$. At level $k=1$ and $k=2$, the classical
moduli space ${\cal M}$  coincides with the infra-red moduli space of $SO(4)$ and
$SO(5)$ super Yang-Mills theory respectively.   For higher Chern-Simons level, the moduli space is a quotient of ${\cal M}$. At a generic point in the
moduli space, the massive spectrum is proportional to the area of
the triangle formed by the two membranes and the orbifold fixed
point.
\end{abstract}

\end{titlepage}

\subsection*{Introduction}

In \cite{bl2}, a novel, conformally invariant, Lagrangian in $d=2+1$
dimensions was constructed. The theory enjoys maximal supersymmetry
and a manifest $SO(8)$ R-symmetry, strongly suggesting that it
describes the low-energy dynamics of multiple M2-branes in M-theory.
Various aspects of this theory were anticipated in \cite{bh,bl,gust}
and  a number of recent papers have explored some of its properties
\cite{bl3,sunil,schwarz,berman,mvr}. Yet so far the interpretation
in terms of M2-branes has remained somewhat murky. The purpose of
this short note is to shed some light on this issue through a study
of the classical vacuum moduli space and spectrum of the theory.

\para
We work with the simplest -- and, to date, only -- explicit
example of the Lagrangian, which is based on an $SO(4)$ gauge
symmetry with an integer valued coupling constant $k$. We will
show that, at levels $k=1$ and $k=2$, the classical moduli space ${\cal M}$
coincides with the infra-red limit of  $SO(4)$ and $SO(5)$ super Yang-Mills theory.
This describes two membranes moving in the background of the orbifold
 ${\bf R}^8/{\bf Z}_2$, without and with discrete torsion respectively.
For $k>2$, we find that the vacuum moduli space is the quotient of ${\cal M}$.
The group acts on the moduli space, but does not appear to have a natural
action on the underlying spacetime.  We further show that, at a generic point in the moduli
space, the mass of the heavy  states is proportional to the area
of the triangle formed by the two membranes and the fixed point, and make some
comments on the implications of this mass formula.

\subsection*{The M2-Brane Lagrangian}

The Lagrangian presented in \cite{bl2} is built around a 3-algebra
${\cal A}$. This is a vector space with basis $T^a$, $a=1,\ldots,
{\rm dim}\,{\cal A}$, endowed with a trilinear antisymmetric
product,
\be [T^a,T^b,T^c]=f^{abc}_{\ \ \ d}T^d .\ee
The algebra is accompanied by an inner product,
$h^{ab}=\Tr(T^aT^b)$,  with which indices may be raised and lowered.
The structure constants of the algebra are then required to be
totally anti-symmetric, $f^{abcd} = f^{[abcd]}$, and to satisfy the
``fundamental identity"
\be f^{aef}_{\ \ \ g}f^{bcdg}-f^{bef}_{\ \ \ g}f^{acdg}+f^{cef}_{\ \
\ g} f^{abdg}-f^{def}_{\ \ \ g}f^{abcg}=0 .\ee
The matter fields consist of 8 algebra-valued scalar fields
$X^I_a$, $I=1,\ldots,8$, transforming in the ${\bf 8}_v$ of $SO(8)$,
together with algebra-valued spinors $\Psi^a$ transforming in the ${\bf
8}_s$ of $SO(8)$. The theory also includes a non-propagating gauge
field $A_\mu^{ab}$. The dynamics is governed by the Lagrangian,
\be {\cal L} &=& -\frac{1}{2} {\cal D}^\mu X^{Ia}\,{\cal D}_\mu
X^I_a + \frac{i}{2}\bar{\Psi}^a \Gamma^\mu{\cal D}_\mu\Psi_a +
\frac{i}{4}\bar{\Psi}_b\Gamma_{IJ}X^I_cX^J_d\Psi_af^{abcd} \nn\\
&& - V(X) +\frac{1}{2}
\epsilon^{\mu\nu\lambda}\left(f_{abcd}A_{\mu}^{ab}\partial_\nu
A_\lambda^{cd}+\frac{2}{3} f_{cda}^{\ \ \
g}f_{efgb}A_\mu^{ab}A_\nu^{cd}A_\lambda^{ef}\right) ,\label{lag}\ee
where the scalar potential is
\be V(X)=
\frac{1}{12}f_{abcd}\,f_{efgd}\,X^{Ia}X^{Jb}X^{Kc}\,X^{Ie}X^{Jf}X^{Kg}
, \ee
while the covariant derivative is defined by
\be {\cal D}_\mu X^{Ia} = \partial_\mu X^{Ia}+f^{a}_{\
bcd}A_{\mu}^{cd}X^{Ib} . \label{d}\ee
The theory is invariant under 16 supercharges and the gauge symmetry:
\be \delta X^I_a &=& -f_{abcd}\Lambda^{bc}X^{Id}\nn\\
\delta \Psi^a &=& -f^{abcd}\Lambda_{bc}\Psi^d\\ f_{abcd}\,\delta
A_\mu^{ab} &=& f_{abcd} {\cal D}_\mu\Lambda^{ab}\nn .\ee
Presently, the only known, finite-dimensional, representation of a
3-algebra has ${\rm dim}\,{\cal A}=4$ and the gauge field
$A_{\mu}^{ab}$ is valued in $so(4)$. The inner product is  taken to
be $h^{ab}=\delta^{ab}$ while the structure constants are
\cite{bl2}\footnote{We have redefined $k$ by a factor of 2 relative to version
1 of this paper: $k_{\rm old} = \ft12 k_{\rm new}$. This is so that that the
$k_{\rm old}=\ft12$ moduli space, mentioned only briefly in a footnote in v1, is elevated
to the $k_{\rm new}=1$ moduli space, as befits the extended discussion given later in the paper. This redefinition is responsible for the apparent differences in subsequent formulae between v1 and the current version.}
\be f^{abcd} = \frac{2\pi}{k}\,\epsilon^{abcd}\label{f} .\ee
In fact, as shown in \cite{bl3,mvr}, for this choice of structure
constants the 3-algebra theory is not as exotic as it first
appears, for it reduces to a familiar Chern-Simons theory with
gauge fields in the Lie algebra $su(2)+ su(2)$ and matter in
the bi-fundamental representation. The requirement that the theory
is invariant under large gauge transformations imposes the usual
quantization on the Chern-Simons coefficient which simply
reads
\be k \in {\bf Z} .\ee
This differs from the result quoted in \cite{bl3} which, with our normalization, was
$k\in 2{\bf Z}$. The correct normalization in the $SO(4)$ case
can be seen by rewriting
the action in terms of $SU(2)\times SU(2)$ gauge fields and, correcting a small
typo in \cite{mvr}, noting that the coefficient of the CS term is $k/4\pi$.
In the rest of this note we study a few elementary aspects of this $SO(4)$ theory.

\subsection*{The Classical Moduli Space}

We start by examining the vacuum moduli space of the classical theory,
defined as solutions to $V(X)=0$ modulo gauge transformations.
This was previously discussed in \cite{bl3,mvr}. However, in both
analyses, there was no obvious interpretation of the moduli space
in terms of known M-theoretic objects. Here we clarify some points
about the appearance of the dual photon which results in a simple
M2-brane interpretation.
%
%\para
%This map will prove particularly useful in elucidating the vacuum
%moduli space of the theory. To this end, following \cite{mvr}, we
%introduce a $2\times 2$ matrix notation for $X^I$
%%
%\be X^I = \frac{1}{2}\left(X^I_4\,{\bf 1} + X^I_i\sigma^i\right)\label{x22}\ee
%%
%To see action of the gauge group, we define
%the self-dual and anti-self-dual gauge fields $A^\pm$
%%
%\be
%A^\pm_{\mu\,ab} = -\frac{\pi}{k}(A_{\mu\,ab}\pm\ft12 \epsilon_{abcd}A_\mu^{cd})\ee
%%
%The $su(2)_\pm$ valued gauge fields are then defined by
%$A^\pm_\mu=A^\pm_{\mu\,4i}\sigma^i$. In this notation, the
%covariant derivative becomes ${\cal D}X^I = \partial
%X^I+iA^+X^I-iX^IA^-$, and the $X^I$ indeed transforms in the
%bi-fundamental representation of $SU(2)_+\times SU(2)_-$ as
%advertised:
%
%\be X^I \rightarrow g_+\,X^I\,g_-^{-1}\label{gg}\ee
%%
%with $g_\pm\in SU(2)_\pm$.

\para
By a suitable gauge transformation, solutions to $V(X)=0$ may be
written as \cite{bl3}
\be X^I = r_1^IT^1 + r_2^IT^2,\qquad {\it i.e.}\qquad  X^I =
\left(\matrix{r_1^I\cr r^I_2\cr0\cr 0\cr}\right)\ .
\label{Xvev}\ee
However, as stressed in \cite{mvr}, there are additional
gauge symmetries which preserve the form of $X^I$ but act
non-trivially on the two eight-dimensional vectors $r_1$ and
$r_2$. Since $X$ transforms in the fundamental representation of
$SO(4)$, we may act by $g\in SO(4)$ in the block diagonal form
\be g = \left(\matrix{g_1&0\cr0&g_2\cr}\right), \label{g}\ee
where $g_1, g_2 \in O(2)$, with $\det g_1 = \det g_2$.
Let us first look at a number of discrete symmetries. Since $g_2$
acts trivially on (\ref{Xvev}) we can effectively ignore it and
simply look at $g_1\in O(2,{\bf Z})$. There are three choices for
$g_1$ which generate all of $O(2,{\bf Z})$ and act on $r_1$ and
$r_2$ as
\be  &\left(\matrix{-1&0\cr0&1\cr}\right)&:\quad r_1 \rightarrow
-r_1\ \ \ ,\ \ \ r_2\rightarrow r_2 \nn\\
&\left(\matrix{1&0\cr0&-1\cr}\right)&: \quad r_1 \rightarrow
r_1\ \ \ ,\ \ \ r_2\rightarrow -r_2 \label{3gs} \\
&\left(\matrix{0&1\cr1&0\cr}\right)&: \quad r_1 \rightarrow r_2\ \
\ ,\ \ \ r_2\rightarrow r_1 .\nn\ee
After imposing these discrete symmetries, $r_1$ and $r_2$
parameterize the 16-dimensional moduli space ${\cal M}\cong (({\bf
R}^8/{\bf Z}_2)\times({\bf R}^8/{\bf Z}_2))/{\bf Z}_2$. However, we
have still to divide out by the continuous $g_1 \in SO(2)\cong
U(1)_{12}$ symmetry, which acts as
\be U(1)_{12}:\ \ z^I \rightarrow e^{i\theta} z^I\ \ \ \ {\rm
where}\ \ \ z^I = r^I_1 + ir^I_2 .\label{u12}\ee
If we make use of all three discrete gauge symmetries \eqn{3gs}, we
already have the identification $z^I\rightarrow iz^I$. Thus, in
order not to overreact, we must take the parameter $\theta$ to
have range $\theta\in[0,\pi/2)$. Alternatively, we could impose just one discrete
symmetry, say the last one which reads $z\rightarrow i\bar{z}$, and take
$\theta\in[0,\pi]$.

\para
Dividing out by this continuous gauge symmetry would seem to leave
us with a 15-dimensional moduli space. This is a rather odd state
of affairs and would contradict the expectations of
supersymmetry. We will now show that by considering the unbroken
gauge symmetry of the theory we will recover this lost dimension
of moduli space. To see this, we proceed by writing down the
low-energy effective action.

\para
Because of the $\epsilon^{abcd}$ appearing in the covariant
derivative \eqn{d}, the $U(1)_{12}$ gauge symmetry \eqn{u12} is
associated to the gauge field $A_\mu^{34}$. Normalizing so that
$z^I$ has charge $+1$, we define
\be B_\mu = \frac{4\pi}{k} A_\mu^{34}\ .\ee
Then the kinetic terms on moduli space are given by
\be {\cal L}_{\rm moduli} = -\frac{1}{2} |{\cal D}_\mu z^I|^2\
.\label{z}\ee
with ${\cal D}z= \partial z +iBz$. At a generic point in moduli
space, there is also an unbroken $SO(2)$ symmetry \cite{mvr},
arising from the action $g_2$ in \eqn{g}. We will call this
symmetry $U(1)_{34}$. It is associated to the gauge field
\be C_\mu = \frac{4\pi}{k} A_\mu^{12}\ ,\ee
where the normalization is again taken to ensure that charged
fields have charge $\pm 1$ under $C_\mu$. (Of course, by
definition the moduli $z^I$ themselves have charge zero under the
unbroken symmetry). A mixed Chern-Simons term couples the $B$ and
$C$ gauge fields;
\be {\cal L}_{cs} = \frac{k}{2\pi}\, \epsilon^{\mu\nu\lambda}\,
B_\mu\partial_\nu C_\lambda \label{cs} .\ee
It was shown in \cite{sunil} that integrating out the broken gauge
field $B$ induces a Maxwell term for $C$, promoting it to a
dynamical field. (In fact, the calculation in \cite{sunil} was
done at a non-generic point in moduli space with an unbroken
$SU(2)$ gauge symmetry, but it proceeds in the same manner  at a
generic point). Here we instead replace the unbroken gauge field
$C$ with its dual photon, introduced in its usual guise as a
Lagrange multiplier to impose the Bianchi identity on the field
strength $G_{\mu\nu}=\partial_\mu C_\nu-\partial_\mu C_\nu$.
\be {\cal L}_{\rm dual} =
-\frac{1}{8\pi}\,\sigma\,\epsilon^{\mu\nu\lambda}\, \partial_\mu
G_{\nu\lambda}\label{dual} .\ee
The normalization is chosen such that $\sigma\in [0,2\pi)$. To see
this, note that $U(1)_{34} \subset SU(2)_{\rm diag} \subset
SO(4)$, with all matter fields in our theory living in the adjoint
of $SU(2)_{\rm diag}$. The magnetic configurations of the theory
are therefore given by the familiar Euclidean 't Hooft-Polyakov
monopole solutions which satisfy the quantization condition,
\be \frac{1}{8\pi} \int d^3x\ \epsilon^{\mu\nu\lambda}
\partial_\mu G_{\nu\lambda}\ \in {\bf Z} , \label{dirac}\ee
%
%
%
%
%Since this periodicity is crucial in what follows, let us take a
%moment to review where it comes from. The usual Dirac quantization
%condition requires that $\int \!\!{}*dG \in 4\pi{\bf Z}$, which
%would suggest that $\sigma$ has periodicity $4\pi$. But this is
%not correct. To understand why this is so, we must examine the
%genealogy of our Abelian subgroup:
%
%. It is well known that the 't Hooft-Polyakov monopoles in such
%theories fails to saturate the Dirac quantization condition, but
%instead gives
%
%\footnote{There are a number of simple ways to see
%that this must be the case. For example, we could choose to add
%further matter in the fundamental representation of $SU(2)_{\rm
%diag}$ which would have charge $+1/2$ under $U(1)_{34}$. This
%matter could be arbitrarily massive, but any monopole
%configuration must still obey the correct Dirac quantization
%\eqn{dirac}. A equivalent explanation, in terms of a ${\bf Z}_2$
%gauge action on $\sigma$, was given in \cite{seiberg}.}
%
%
In the presence of the mixed Chern-Simons term \eqn{cs}, the shift
symmetry of the dual photon becomes gauged under $U(1)_{12}$. This
follows because the topological current
${}^*G$, which generates the shift symmetry of the dual photon, is
coupled to $B_\mu$. It is also simple to see by collecting
together the various pieces of the Lagrangian, which can be found
in \eqn{z}, \eqn{cs} and \eqn{dual},
\be {\cal L}_{\rm moduli}+{\cal L}_{CS} +{\cal L}_{\rm dual} =
-\frac{1}{2}|{\cal D}_\mu z^I|^2 +
\frac{1}{8\pi}\,\epsilon^{\mu\nu\lambda}\,\left(2k B_\mu
+\partial_\mu\sigma\right)G_{\nu\lambda} .\label{19}\ee
This is invariant under the gauge action
\be U(1)_{12}:\ \ \ z^I\rightarrow e^{i\theta}z^I\ \ \ ,\ \ \ \sigma
\rightarrow \sigma + 2k \,\theta\label{gt}\ \ \ ,\ \ \ B_\mu
\rightarrow B_\mu - \partial_\mu\theta .\ee
Together with the discrete gauge symmetries \eqn{3gs}, which now
also induce a sign flip for $\sigma$.
%
%\be \Omega&:& z^I\rightarrow i(z^I)^{\dagger} \ \ ,\ \
%\sigma \rightarrow 2\pi-\sigma\nn \\ %\Theta&:& z^I\rightarrow (z^I)^{\dagger} \ \
%,\ \ \sigma \rightarrow 2\pi-\sigma
%\ee
%

\para
In the next section, we will use \eqn{19} to analyze the moduli space dynamics.
However, we can go further and eliminate the field strength
$G_{\mu\nu}$. Since it is now unconstrained by the Bianchi identity,
it acts as a Lagrange multiplier imposing the requirement that
$B_\mu = -(1/2k)
\partial_\mu\sigma$ is pure gauge. This results in the action
\be {\cal L}= -\frac{1}{2}|\partial_\mu z^I - \frac{i}{2k} z^I
\partial_\mu\sigma |^2 ,\ee
and we observe that $\sigma$ can be eliminated by the field
redefinition $z^I\to e^{-i\sigma/2k}z^I$. However, this
transformation still leaves us with a number of discrete
identifications which we now examine more carefully.

\subsection*{The Theory at Level $k=1$ and $k=2$}

Let us return to the action in the form \eqn{19}. For $k=1$, we impose just one of the discrete symmetries, which we take to be $z\rightarrow i\bar{z}$, with $\theta\in[0,\pi]$. We can now fix the $U(1)_{12}$ gauge symmetry by imposing $\sigma=0$, leaving us with remnant
${\bf Z}_2$ which acts by $\sigma\rightarrow \sigma+2\pi$ and $z\rightarrow -z$. The moduli space at level $k=1$ is thus,
\be {\cal M}_{k=1} \cong \frac{{\bf R}^8\times {\bf R}^8}{\bf{Z}_2\times {\bf Z}_2}
\label{m12}\ee
Writing $z=r_1+ir_2$, the two ${\bf Z}_2$ factors act as $(r_1,r_2)\rightarrow (-r_1,-r_2)$ and $(r_1,r_2)\rightarrow (r_2,r_1)$. As observed in \cite{rival}, this
coincides with the infra-red limit of the moduli space of $d=2+1$ dimensional, maximally supersymmetric Yang-Mills (SYM) with $SO(4)$ gauge group.

\para
For $k=2$, we
may again fix the $U(1)_{12}$ gauge symmetry by setting $\sigma=0$. Imposing
all three discrete symmetries, we have $\theta\in [0,\pi/2)$ which now leaves no
further
residual transformation. The moduli space dynamics is simply given
by the 8 complex scalars $z^I$, endowed with a flat metric and subject
to the  discrete symmetries \eqn{3gs}. We
conclude that the classical vacuum moduli space of the theory at
level $k=2$ is
\be {\cal M}_{k=2} \cong \frac{({\bf R}^8/{\bf Z_2}) \times ({\bf
R}^8/{\bf Z}_2)}{{\bf Z}_2} .\label{m1}\ee
The coincides with the moduli space of $SO(5)$ SYM in the infra-red limit or, alternatively,
the configuration space of two M2-branes in
the background of the orbifold ${\bf R}^8/{\bf Z}_2$.

\para
We strike a note of caution: the $k=1$ and $k=2$ theories are strongly coupled at
all points in their moduli space. Nonetheless, we will assume that we can
take \eqn{m12} and \eqn{m1} at face value. We take this as evidence that the $k=1$ and
$k=2$ theories describe the infra-red fixed point of $SO(4)$ and $SO(5)$ SYM respectively \footnote{This interpretation differs from that offered in
\cite{bl3,sunil}. In particular, in \cite{bl3}, $r_1$ and $r_2$
were viewed as the relative separation of 3 M2-branes. However,
neither the discrete symmetries, nor the flat diagonal metric,
lend support to this.}. As we now review, in each case
this can be understood as M2-branes moving
in the orbifold background ${\bf R}^8/{\bf Z}_2$.

\para
Let us briefly review a few pertinent facts about the M-theory
orbifold  ${\bf R}^8/{\bf Z}_2$. There are actually two different
such orbifolds, distinguished by discrete torsion for $G_4$
arising because $H^4({\bf RP}^7,{\bf Z}) \cong {\bf Z}_2$
\cite{sethi}. The orbifolds  with and without torsion are referred
to as type-B and type-A  respectively. The low-energy dynamics of
$N$ M2-branes in these orbifold backgrounds is thought to be
governed by a maximally supersymmetric,  $SO(8)$ invariant
conformal fixed point. These arise as the strong coupling limit of
maximally supersymmetric Yang-Mills (SYM) in $d=2+1$ dimensions
with gauge groups $O(2N)$, $SO(2N+1)$, and $Sp(N)$. As explained
in \cite{sethi,bk}, the fact that these three classical groups
flow to one of only two possible theories  implies non-trivial IR
dualities  between distinct UV theories. The RG flows occur as
follows: $O(2N)$ SYM flows to the theory on M2-branes on the
A-type orbifold; $SO(2N+1)$ SYM flows to the theory on the B-type
orbifold; while $Sp(N)$ SYM flows to either the theory on the
A-type or B-type orbifold, depending on the expectation value of
the dual photon. Comparing to our previous analysis, we see that the $k=1$ theory
describes two membranes on the A-type orbifold, while the $k=2$ theory
describes two membranes on the B-type orbifold.

\para
The identification of the M2-brane Lagrangian \eqn{lag} with M2-branes on
an orbifold also resolves a puzzle raised in \cite{mvr} regarding
chiral primary operators. The bosonic, gauge invariant, operators
of \eqn{lag} live in tensor representations of SO(8) with
an even number of indices. Yet the chiral primary operators
derived from M-theory on $AdS_4 \times {\bf S}^7$ live in the
symmetric traceless $s$-index representations of SO(8), with both
even and odd $s$. However, pleasingly only the even $s$
representations survive the orbifold projection in supergravity
\cite{aoy}. Although the AdS/CFT analysis is valid only at large
$N$, it is comforting that this basic feature agrees with the
$N=2$ M2-brane theory.

\subsection*{The Theory at Level $k>2$}

Perhaps the most intriguing consequence of  the
Lagrangian \eqn{lag} is the
existence of a weakly coupled limit when $k\gg 1$.
Understanding how such a limit arises from an
M-theoretic description may be our best hope of getting a handle
on the underlying microscopic degrees of freedom.

\para
For $k>2$, setting $\sigma =0$ does not completely fix the
$U(1)_{12}$ gauge action \eqn{gt}.  There exists a residual ${\bf
Z}_k$ symmetry which leaves $\sigma =0\ {\rm mod}\ 2\pi$ and is
generated by,
\be  z^I \rightarrow e^{i \pi/ k} z^{I} .\label{zk}\ee
As pointed out in \cite{rival}, this ${\bf Z}_k$ action does not commute
with the ${\bf Z}_2$ actions of equation \eqn{3gs}. Between them they
generate the dihedral group $D_{2k}$. We conclude that the moduli space
is given by,
\be {\cal M}_k \cong \frac{{\bf R}^8\times {\bf R}^8}{D_{2k}}\label{dk}\ee
However, while the group $D_{2k}$ has a simple action on the moduli
space, it does not appear to have a such a description on the
spacetime transverse to the M2-branes for $k>2$. In particular, it does not
leave the distances between branes fixed. Needless to say, it would
be potentially rather interesting to better understand the microscopic meaning
of this quotient action and these higher $k$ theories. A curious observation
of \cite{rival} is that the moduli space for $k=3$ coincides with the infra-red
limit of SYM with $G_2$ gauge group.

\subsection*{The Spectrum and Non-Abelian Gauge Restoration}

We note that the ${\bf Z}_k$ action \eqn{zk} would not make much
sense on a pair of D-branes.  One simple way to see this is to
note that it does not preserve the distance between the two
branes. In string theory this distance dictates the spectrum of
massive states arising from stretched strings. Yet the M2-brane
theory appears to be blind to the transverse distance between the
two branes. It knows only about transverse areas! This is clear if we
look at the classical mass spectrum, which we trust for $k\gg
1$. Sitting at a generic point in moduli space, we may employ the
$SO(8)$ R-symmetry to rotate the M2-branes to lie in the $X^7-X^8$
plane. Then the mass of states is given by,
\be M = \frac{4\pi}{k} A\label{ma}\ee
where $A=\ft12|r_1^7 r_2^8 - r_1^8r_2^7|=\ft14|\bar{z}^7z^8 -
\bar{z}^8z^7|$ is the area of the triangle formed by the two
M2-branes and the orbifold fixed point. This is manifestly
invariant under the ${\bf Z}_k$ action.

\para
We finish with a few comments on the implications of this mass formula. Firstly, it implies that new states become massless when the
branes become co-linear with the orbifold fixed point. This is to be contrasted
with the familiar statement that states on D-branes become massless when branes
coincide. Let us see how these massless states arise.
In generic vacua the R-symmetry is broken to
$SO(6)$ and, as we noted previously, a $U(1)_{34}$ gauge symmetry survives.
However, when the branes are
co-linear, and the R-symmetry is broken to $SO(7)$, a full $SO(3)$ gauge symmetry
is left unbroken. This was the situation examined in \cite{sunil} where it
was shown that, upon integrating out the broken gauge generators, this $SO(3)$
gauge field becomes dynamical. These are the new massless states.

\para
The emergence of this dynamical $SO(3)$ gauge field
is something of a blessing, for it removes a potential difficulty in interpreting
the expectation value \eqn{Xvev} as the position of two branes. The problem is that
whenever the branes are co-linear, one can change the relative positions of the branes
through a gauge transformation. For example, the $SO(7)$ preserving expectation values
\be X^I = r^I(c_1T^1+c_2T^2)\label{cvac}\ee
are gauge equivalent for all $c_1$ and $c_2$ such that $c_1^2+c_2^2$ is constant. Naively this would equate configurations with different separations between co-linear branes and the fixed point. In fact the theory does distinguish between these configurations, but it is somewhat hard to see explicitly. The presence of the
dynamical, unbroken, $SO(3)$ gauge field means that there is a non-Abelian dual photon, whose expectation value will determine the relative positions of the branes. This is entirely analogous to the situation of two D2-branes in IIA string theory, for which the moduli space is $({\bf R}^7\times {\bf S}^1)/{\bf Z}_2$. Even at the origin of ${\bf R}^7$, where the gauge group is unbroken, the branes may still be separated in a non-singular fashion along the M-theory circle. However, seeing this how this explicitly arises from the non-Abelian dual photon is difficult.

\para
A related fact is that the appearance of the massless states when the branes are
co-linear does not necessarily imply a singularity in the low-energy effective theory.
This is exemplified in the D2-brane , where there are only isolated singularities in the moduli space, rather than a whole ${\bf S}^1$'s worth of singularities at the origin of
${\bf R}^7$. Indeed, from the M-theory perspective, the generic point with co-linear branes should be smooth. More precisely, we expect that, in the vacua \eqn{cvac}, there is just a single singularity for the $k=1$ theory, corresponding to the two
two branes sitting on top of each other. For the $k=2$ theory, there should be two singularities, the first corresponding to the two branes sitting on top of each other, while the second corresponds to one brane sitting on the orbifold fixed plane which is now expected to result in a non-trivial fixed point.

\para
Finally, it is tempting to believe that the mass formula \eqn{ma} is hinting at
some fundamental degree of freedom of M-theory. The fact that the mass should scale
as an area is, for $k\gg 1$,  a consequence of conformal invariance,
and the triangle is the only natural area in the theory.
Nonetheless, the appearance of such a ``3-pronged" object is
intriguing, not least because such states would naively explain
the famous $N^3$ entropy of the M5-brane theory \cite{igor}.
However, quite
how one could scale such states to account for the $N^{3/2}$
entropy for M2-branes, in a controllable weakly coupled regime,
appears as tantalisingly mysterious as ever.

\subsection*{Acknowledgments}

We would like to thank Nick Dorey, Ami Hanany, Rajesh Gopakumar and Julian Sonner for
useful discussions. We are especially grateful to Jacques Distler, Sunil Mukhi,
Costis Papageorgakis and Mark van Raamsdonk for detailed and enjoyable conversations to iron  out small factor of 2 discrepancies between version 1 of
\cite{rival} and version 1 of this paper.   NL is supported in part by the
PPARC grant PP/C507145/1 and the EU grant MRTN-CT-2004-512194. DT is
supported by the Royal Society.


\begin{thebibliography}{99}

\small
\parskip=0pt plus 2pt


\bibitem{bl2}
  J.~Bagger and N.~Lambert,
  ``{\it Gauge Symmetry and Supersymmetry of Multiple M2-Branes},''
  Phys.\ Rev.\  D {\bf 77}, 065008 (2008)
  [arXiv:0711.0955 [hep-th]].
  %%CITATION = PHRVA,D77,065008;%%



\bibitem{bh}   A.~Basu and J.~A.~Harvey,
  ``{\it The M2-M5 brane system and a generalized Nahm's equation},''
  Nucl.\ Phys.\  B {\bf 713}, 136 (2005)
  [arXiv:hep-th/0412310].
  %%CITATION = NUPHA,B713,136;%%


\bibitem{bl}
  J.~Bagger and N.~Lambert,
  ``{\it Modeling multiple M2's},''
  Phys.\ Rev.\  D {\bf 75}, 045020 (2007)
  [arXiv:hep-th/0611108].
  %%CITATION = PHRVA,D75,045020;%%

\bibitem{gust}
  A.~Gustavsson,
  ``{\it Algebraic structures on parallel M2-branes},''
  arXiv:0709.1260 [hep-th].
  %%CITATION = ARXIV:0709.1260;%%

\bibitem{bl3}
  J.~Bagger and N.~Lambert,
  ``{\it Comments On Multiple M2-branes},''
  JHEP {\bf 0802}, 105 (2008)
  [arXiv:0712.3738 [hep-th]].
  %%CITATION = JHEPA,0802,105;%%


\bibitem{sunil}  S.~Mukhi and C.~Papageorgakis, ``{\it M2 to D2},''
  arXiv:0803.3218 [hep-th].
  %%CITATION = ARXIV:0803.3218;%%

\bibitem{schwarz}   M.~A.~Bandres, A.~E.~Lipstein and J.~H.~Schwarz,
  ``{\it $N=8$ Superconformal Chern--Simons Theories},''
  arXiv:0803.3242 [hep-th].
  %%CITATION = ARXIV:0803.3242;%%

\bibitem{mvr}
  M.~Van Raamsdonk,
  ``{\it Comments on the Bagger-Lambert theory and multiple M2-branes},''
  arXiv:0803.3803 [hep-th].
  %%CITATION = ARXIV:0803.3803;%%

\bibitem{berman}
  D.~S.~Berman, L.~C.~Tadrowski and D.~C.~Thompson,
  ``{\it Aspects of Multiple Membranes},''
  arXiv:0803.3611 [hep-th].
  %%CITATION = ARXIV:0803.3611;%%


\bibitem{sethi}   S.~Sethi,
  ``{\it A relation between N = 8 gauge theories in three dimensions},''
  JHEP {\bf 9811}, 003 (1998)
  [arXiv:hep-th/9809162].
  %%CITATION = JHEPA,9811,003;%%

\bibitem{bk}   M.~Berkooz and A.~Kapustin,
  ``{\it New IR dualities in supersymmetric gauge theory in three dimensions},''
  JHEP {\bf 9902}, 009 (1999)
  [arXiv:hep-th/9810257].
  %%CITATION = JHEPA,9902,009;%%

\bibitem{aoy}   O.~Aharony, Y.~Oz and Z.~Yin,
  ``{\it M-theory on $AdS_p\times S^{11-p}$ and superconformal field theories},''
  Phys.\ Lett.\  B {\bf 430}, 87 (1998)
  [arXiv:hep-th/9803051].
  %%CITATION = PHLTA,B430,87;%%

\bibitem{rival}   J.~Distler, S.~Mukhi, C.~Papageorgakis and M.~Van Raamsdonk,
  ``{\it M2-branes on M-folds},''
  arXiv:0804.1256 [hep-th].
  %%CITATION = ARXIV:0804.1256;%%

\bibitem{igor}  I.~R.~Klebanov and A.~A.~Tseytlin,
  ``{\it Entropy of Near-Extremal Black p-branes},''
  Nucl.\ Phys.\  B {\bf 475}, 164 (1996)
  [arXiv:hep-th/9604089].
  %%CITATION = NUPHA,B475,164;%%

\end{thebibliography}
\end{document}